\documentclass{article}

\usepackage{PRIMEarxiv}
\usepackage{amsmath}
\usepackage[utf8]{inputenc} 
\usepackage[T1]{fontenc}    
\usepackage{hyperref}       
\usepackage{url}            
\usepackage{booktabs}       
\usepackage{amsfonts}       
\usepackage{nicefrac}       
\usepackage{microtype}      
\usepackage{lipsum}
\usepackage{fancyhdr}       
\usepackage{graphicx}       
\graphicspath{{media/}}     

\title{Segmentation of Mental Foramen in Orthopantomographs: A Deep Learning Approach
}

\author{
  Haider Raza, Mohsin Ali, Vishal Krishna Singh \\
  School of Computer Science and Electronic Engineering, \\
  University of Essex,\\
  Colchester, UK. \\
   \And
  Agustin Wahjuningrum \\
  Department of Conservative Dentistry,  \\
  Faculty of Dental Medicine, University of Airlangga, \\
  Surabaya, Indonesia.\\
  \AND
  Rachel Sarig \\
  Department of Oral Biology, \\
  Tel Aviv University, 6997801, \\
  Tel Aviv, Israel. \\
  \And
  Akhilanand Chaurasia \\
  Department of Oral Medicine and Radiology, \\
  King George's Medical University,\\
  Lucknow, India. \\
}

\begin{document}
\maketitle

\begin{abstract}
Precise identification and detection of the Mental Foramen are crucial in dentistry, impacting procedures such as impacted tooth removal, cyst surgeries, and implants. Accurately identifying this anatomical feature is important for facilitating post-surgery issues and improving patient outcomes. Moreover, this study aims to accelerate dental procedures, elevating patient care and healthcare efficiency in dentistry. This research used Deep Learning methods to accurately detect and segment the Mental Foramen from panoramic radiograph images. Two mask types, circular and square, were used during model training. Multiple segmentation models were employed to identify and segment the Mental Foramen, and their effectiveness was evaluated using diverse metrics. An in-house dataset comprising 1000 panoramic radiographs was created for this study. Our experiments demonstrated that the Classical U–Net model performed exceptionally well on the test data, achieving a Dice Coefficient of 0.79 and an Intersection over Union (IoU) of 0.67. Moreover, ResUNet++ and U–Net Attention models showed competitive performance, with Dice scores of 0.675 and 0.676, and IoU values of 0.683 and 0.671, respectively. We also investigated transfer learning models with varied backbone architectures, finding LinkNet to produce the best outcomes. In conclusion, our research highlights the efficacy of the classical U–net model in accurately identifying and outlining the Mental Foramen in panoramic radiographs. While vital, this task is comparatively simpler than segmenting complex medical datasets such as brain tumours or skin cancer, given their diverse sizes and shapes. This research also holds value in optimizing dental practice, benefiting practitioners and patients alike.
\end{abstract}


\section{Introduction}
Medical imaging, encompassing modalities such as computed tomography (CT), cone-beam computed tomography (CBCT), magnetic resonance imaging (MRI), positron emission tomography (PET), mammography, ultrasound, X-ray, and more, has played a pivotal role in early disease detection, diagnosis, and treatment over the past few decades \cite{naumova2014clinical}. The generation of vast digital data and the inherent variability in pathologies across different populations have prompted the integration of computer-based approaches by researchers and medical professionals. While medical imaging has seen significant advancements, computational medical image analysis has sometimes lagged behind. Recent strides in deep learning and computer vision, however, have led to transformative breakthroughs, including the use of deep learning for tasks such as COVID-19 detection \cite{saood2021covid} and forensic dentistry \cite{wang2023population}. Traditionally, feature preparation, selection, and extraction in machine learning heavily relied on expert knowledge and domain-specific insights \cite{janowczyk2016deep}. Deep learning has revolutionized this paradigm by automating feature extraction, empowering machines to perform tasks that were once the domain of human experts \cite{chen2016deep}. These advancements have been facilitated by parallel developments in hardware technology, including high-speed Central Processing Units (CPUs) and Graphics Processing Units (GPUs), as well as the availability of extensive datasets and the evolution of learning algorithms. Consequently, deep learning-based computer vision techniques have been successfully applied to analyze medical images, detect diseases, and precisely locate anatomical regions requiring medical intervention \cite{puttagunta2021medical}. Among the critical tasks in medical image analysis, image segmentation stands out. It is pivotal in numerous medical domains, allowing the localization of regions of interest (ROIs) within scanned images like MRI, CT, or X-rays. Convolutional Neural Networks (CNNs) have emerged as effective tools for this purpose.

From a dentistry perspective, the mental foramen (MF) holds significant clinical importance. Located on the buccal surface of the mandibular cortex, the MF serves as an anatomical landmark for dental treatments such as local anaesthesia injections in maxillofacial and plastic surgery \cite{laher2016finding}. It houses the mental nerve, the terminal branch of the inferior alveolar nerve, and associated blood vessels. Dental procedures involving the MF carry the risk of complications, including hematomas, anaesthetic toxicity, and neurosensory disturbances \cite{smith2006nerve}. While the MF is often visualized in 3D images like CBCT and CT, 2D images, such as Orthopantomographs (OPG), are common in dental clinics. However, interpreting radiographic MF images requires precision. Studies have revealed that some MFs may not be as conspicuous as assumed on OPGs, posing a challenge to clinicians \cite{jacobs2004appearance, yosue1989appearance}. An inexperienced clinician might misinterpret MF localization, highlighting the need for advanced and automated systems. In the context of computer vision applications in dentistry, various CNN architectures have been proposed. Initial approaches, like fully connected convolutional neural networks (fCNN), produced lower-resolution images due to the consecutive use of convolution and pooling layers. To address this limitation, a 3-layer convolutional encoder network was introduced for higher-resolution segmentation. More recently, the U-Net architecture, with its combination of convolution and deconvolution layers and skip connections, has demonstrated the ability to generate highly accurate segmentation probability maps \cite{ronneberger2015u}.

Residual Networks (ResNet) tackled the challenge of vanishing gradients in deep networks by introducing skip connections. These connections enable the learning of residual mappings, alleviating gradient issues and potentially allowing the training of deeper UNet variants \cite{targ2016resnet}. Attention U-Net incorporates attention blocks, which generate attention maps indicating the importance of each pixel's features within the entire image. This selective amplification of informative regions enables the model to focus on critical image areas while suppressing irrelevant information. Multitasking, involving segmentation, classification, and regression, can enhance segmentation precision.

In this paper, we introduce a novel approach to MF segmentation by employing round and square-shaped masks for training image segmentation models. While round masks have been used in previous work, square-shaped masks as ground truth for MF detection and segmentation are introduced. Our study evaluates various architecture variants based on U-Net, assessing their performance using a range of metrics, including dice coefficient score (DSC), and intersection over union (IoU) on validation datasets. Additionally, we report their performance on test datasets, primarily focusing on DSC and IoU metrics. Our investigation begins with assessing U-Net, a cornerstone architecture for medical image segmentation, followed by evaluating advanced models such as ResUNet, Attention U-Net, LinkNet, and Feature Pyramid Network for MF segmentation tasks.

\section{Material and methods}
This paper focuses on improving the accuracy of Mental Foramen (MF) detection and segmentation using a fully connected convolutional neural network (FCNN). To achieve this goal, we employed a range of neural network architectures, including U-Net, U-Net++, ResUNet, ResUNet++, U-Net Attention, DeepLabV3, V-Net, R2Net, FPN (Feature Pyramid Network), and LinkNet. Our study aims to enhance the precision of MF detection and segmentation, addressing specific challenges in this critical task.

\subsection{Material}

\subsubsection*{U--Net} 

The initial neural network model employed for the segmentation of the Mental Foramen (MF) is based on the U-Net architecture. U-Net comprises two primary segments: the contraction (down-sampling) and the expansion (up-sampling) phases \cite{oktay2018attention}. In the contraction path, 3x3 convolutional neural network (CNN) layers with the rectified linear unit (ReLU) activation functions and same padding are used. Additionally, dropout rates gradually increase from 0.1 to 0.5 across the layers. Weight initialization employs the normal method, accompanied by 2x2 max-pooling layers. Conversely, the expansion path incorporates 3x3 convolutional layers and a 4x4 transposed convolutional layer with a stride of 2. It also features concatenation between the contraction and expansion layers to integrate additional information. The final convolutional layer in this path utilizes the sigmoid function to generate the heuristic segmentation map, with values ranging from 0 to 1. However, ReLU activation functions are employed for the remaining CNN layers.

\subsubsection*{U--Net++}
U-Net++ is an extension of the U-Net architecture introduced by Zhou et al. in 2018 \cite{zhou2018unet++}, aimed at enhancing the segmentation performance. It incorporates three key modifications to improve outcomes. Firstly, U-Net++ addresses the semantic gap between the encoder and decoder by incorporating skip pathways on convolutional neural network (CNN) layers. This facilitates the flow of information and reduces the vanishing gradient problem. Secondly, it introduces skip pathways on dense layers, further mitigating the vanishing gradient issue and improving gradient flow. Lastly, U-Net++ enables pruning in deep supervision, which helps to refine the loss function and enhance the overall segmentation quality. These changes collectively contribute to the improved performance and effectiveness of U-Net++ in semantic segmentation tasks.

\subsubsection*{ResUNet}
The Residual U-Net (ResUNet) architecture incorporates residual blocks into the U-Net model to achieve high performance while minimizing the number of parameters required \cite{zhang2018road}. It serves as an advanced iteration of the original U-Net. After each layer, the residual block seeks to learn the residual output by directly adding the input to the subsequent input layer and subtracting the output from the input layer. This approach expedites forward propagation through the inclusion of shortcut layers. The result is increased efficiency and effectiveness of the architecture by allowing for faster information flow and alleviating the vanishing gradient problem.

\subsubsection*{Attention U--Net}
The Attention U-Net architecture in image segmentation focuses exclusively on relevant features during training. This approach enables the network to improve its learning process by conserving computational resources that would otherwise be expended on irrelevant features. Two distinct types of attention mechanisms exist: hard and soft attention \cite{oktay2018attention}. For this study, soft attention is employed, where different regions of the image are assigned varying weights, with greater attention directed toward relevant areas. Consequently, during training, the model prioritizes the essential portions of the image.

\subsubsection*{LinkNet}
LinkNet is characterized by its encoder and decoder blocks, facilitating information transfer after each down-sampling step. This unique design surpasses conventional approaches employing Convolutional Neural Networks (CNNs) or pooling layers in the decoder. LinkNet's approach enhances accuracy by utilizing fewer parameters in the decoder, optimizing computational efficiency. In its initial configuration, the model incorporates a 7x7 kernel size with a stride of 2 in the first layer, followed by a max-pooling layer of size 2 with the same stride size of 2 \cite{chaurasia2017linknet}. This arrangement ensures efficient down-sampling of the input data, a crucial step in various computer vision tasks such as image segmentation.

\subsubsection*{Feature Pyramid Networks (FPN)}
The Feature Pyramid Network (FPN) is a feature extractor designed to process images of arbitrary sizes and generate multi-scale feature maps arranged in a pyramid-like structure. This structure enhances information quality and overall performance \cite{lin2017feature}. FPN consists of two pathways: the bottom-up and the top-down pathways. The bottom-up pathway comprises convolutional layers responsible for extracting features from the input image. As information moves upwards through the layers, spatial resolution decreases while semantic value increases, capturing more abstract and meaningful information. The top-down pathway performs upsampling on lower-resolution layers, using the two closest neighbours to regain finer details and improve spatial resolution \cite{Jonathan:2018}. The combination of the bottom-up and top-down pathways in FPN results in a powerful feature extraction mechanism beneficial for various computer vision tasks, including object detection and segmentation.

\subsection*{Dataset Collection and Workflow}

\begin{figure}[h!]
\centering
\includegraphics[width=12cm, height=6cm]{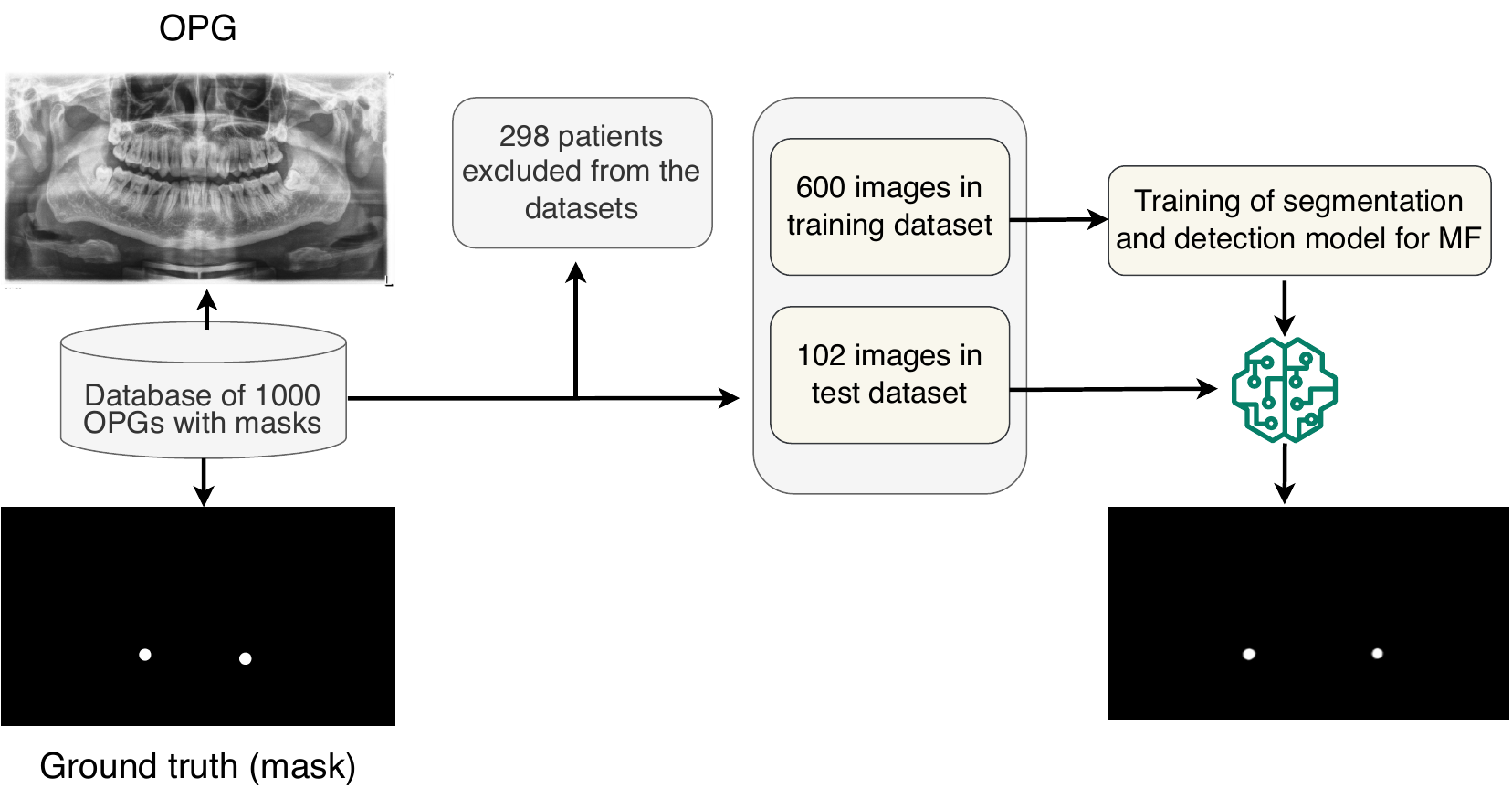} 
\caption{Overall workflow for detecting and segmenting of Mental Foramen in OPGs.}
\label{Fig_Workflow}
\end{figure}

The overall workflow of this study is illustrated in Figure \ref{Fig_Workflow}, which provides a visual representation of the research process. Data collection was carried out meticulously to ensure data consistency and ethical standards. Between December 1, 2021, and December 20, 2022, one thousand Orthopantomogram (OPG) X-ray images were gathered from patients who consecutively visited the Faculty of Dental Medicine, University of Airlangga, Surabaya.
The dataset was specifically curated at the Faculty of Medicine, University of Airlangga, Indonesia to maintain data consistency and accuracy for semantic segmentation.

Ethical approval for the experimental procedures was granted by the University Research Ethics Committee, University of Airlangga (Ethical Clearance Number: 768/HRECC.FODM/X/2022), 
and all research protocols adhered to approved institutional guidelines and regulations, as well as the principles outlined in the Helsinki Declaration. Prior to data acquisition, a comprehensive informed consenting process was followed, wherein all participants were provided with detailed information about the purpose and procedures of the experiments. Written informed consent was obtained from all participants, granting permission for the use of their anonymized data for research purposes by other researchers. Personal identification information was meticulously removed from the dataset to safeguard privacy. Additionally, our research includes images of participants, particularly OPGs, as part of the study involving medical imaging for research and development purposes. We have diligently followed ethical guidelines and privacy protocols throughout our research. We have obtained explicit consent from all participants involved, allowing the usage and publication of their images for research and academic purposes. This consent covers the inclusion of their X-ray images in the research paper and any subsequent publications associated with this work.

The initial dataset comprised 1,000 panoramic radiographic dental images, which were instrumental in investigating the precise location of the mental foramen. During the dataset organization process, careful consideration was given to maintaining a structured and comprehensive approach. The dataset was structured as follows: within one primary folder, three distinct sub-folders were created to house the images. The first sub-folder contained 1,000 raw orthopantomography (OPG) images. These raw images served as the foundation for our study. The second sub-folder comprised 1,000 segmented images, which were meticulously created to delineate the boundaries of the mental foramen. The last sub-folder held 1,000 masked images, where the mental foramen was marked using two distinct masks: round-shaped and square-shaped.

The inclusion of both round and square masks in the dataset was a deliberate choice made during the dataset annotation process. This decision was driven by the uncertainty surrounding which mask shape would yield the most effective results for our specific research objectives. By annotating the dataset with both round and square masks, we aimed to explore and compare the performance of segmentation models when confronted with these variations in mask shape. 

In our research, we recognized the importance of transparency and completeness. Therefore, we opted to report and analyze the results obtained from both the round and square masks. This approach ensures that our findings are comprehensive and provide valuable insights into the segmentation of the mental foramen, allowing us to make informed conclusions based on the performance of different mask shapes.

However, during the annotation of the Orthopantomogram (OPG) X-ray images, our team of dental experts identified certain issues with a subset of the OPGs. These issues included occlusion, noise, or pixel corruption in the proximity of or at the location of the Mental Foramen (MF). These factors rendered some OPGs unsuitable for accurate MF detection. In light of these limitations, we made the decision with consent from dental experts to remove 298 patient images from the original dataset as shown in in Figure \ref{Fig_Workflow}. This curation process was essential to maintain the overall integrity of the dataset and to ensure that it remained suitable for the training and evaluation of deep learning models. Removing these problematic images helped prevent confusion and inaccuracies in the subsequent stages of the study, ultimately contributing to the reliability of our deep learning model's performance.

\subsection{Methods}

\subsection*{Model Training and Data Split}

The models were trained using Orthopantomogram (OPG) radiographs and their corresponding segmentation masks. For this experiment, two different ground truth settings were considered: 1) a round-shaped mask and 2) a square-shaped mask.
After careful data curation, which involved discarding noisy images in consultation with medical experts from the Department of Oral Medicine and Radiology, King George's Medical University, 
the final dataset consisted of 702 radiographs.
From this curated dataset, 600 images were randomly selected for model training. The remaining 102 images were exclusively reserved for testing the performance of the models.

The U-Net architecture and its variants were employed for the training process. The training utilized an Adam optimizer with gradient descent computation and a fixed learning rate of 1e-4. The implementation was carried out using Keras with the TensorFlow backend. Furthermore, early stopping was employed to avoid overfitting of the model. The entire study was conducted and trained on Google Colab, harnessing its inbuilt high-end GPU A-100 for enhanced computational performance.

\subsubsection*{Cross Validation}

\begin{figure}[h!]
\centering
\includegraphics[width=13cm]{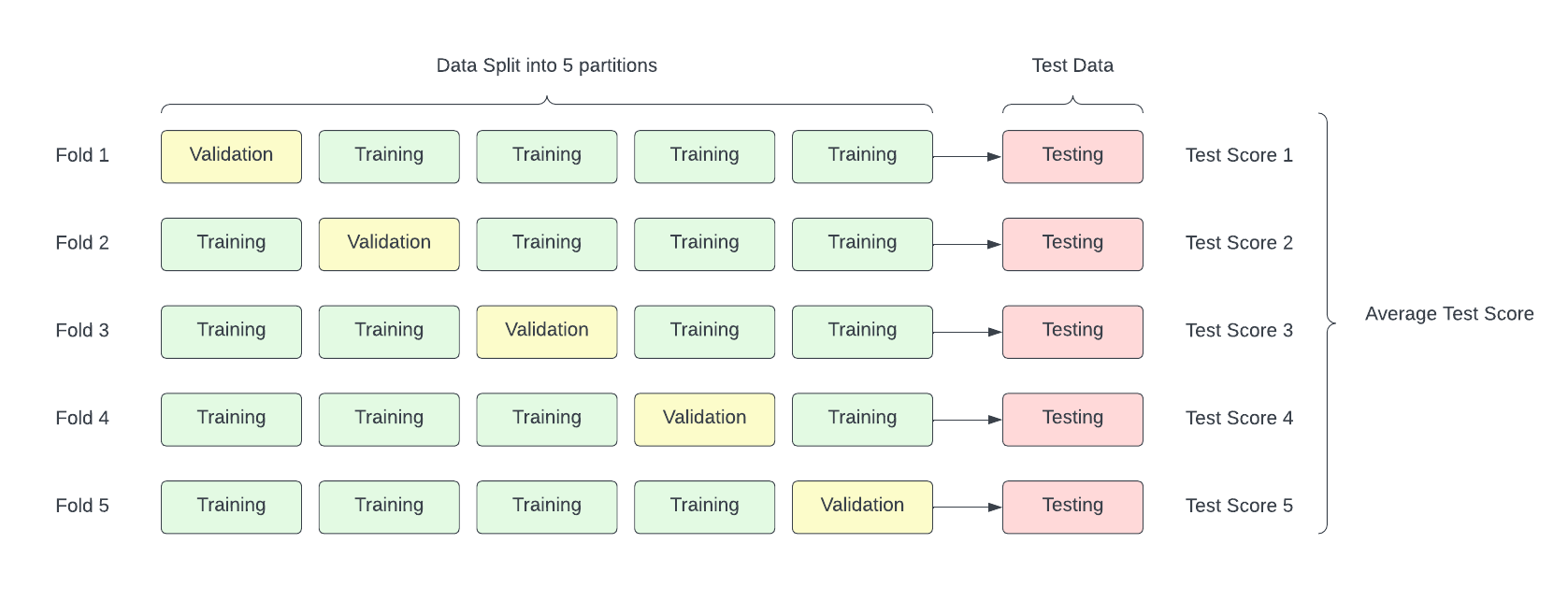}
\caption{5-fold Cross Validation(CV)}
\label{Fig_CV}
\end{figure}

To ensure unbiased and robust results, a 5-fold cross-validation (5-fold CV) methodology was applied to the training dataset, comprising 600 randomly selected radiographs as illustrated in Figure \ref{Fig_CV}. In this approach, the dataset was partitioned into 5 equal subsamples. During each iteration of cross-validation, four of these subsamples were utilized for training the model, while the remaining one was reserved as a validation set for evaluating the model's performance on previously unseen data, with the implementation of early stopping. This process was iteratively repeated until all five subsamples had been employed as the test data exactly once. After each step, test scores was calculated using the 102 testing samples. Subsequently, the test scores for IoU and DSC results obtained from each iteration were averaged to produce a single estimation, which served as the final outcome.

\subsection*{Metrics used for Evaluation}

The performance of the segmentation models was evaluated using the following metrics:



1. \textbf{Dice Similarity Coefficient (DSC)}: The Dice similarity coefficient serves as a pivotal metric for segmentation tasks. It quantifies the degree of overlap between the segmented area and the ground truth. In essence, it measures how well the predicted region aligns with the actual region of interest. This coefficient offers a nuanced assessment by considering both false positives and false negatives. A high DSC value indicates a robust model capable of accurately capturing the target region.

The formula for DSC computes the double intersection of pixels in both the predicted and ground truth regions, normalized by the total number of pixels in the two regions. This normalization accounts for variations in region size, making it a versatile metric suitable for diverse applications. The Dice similarity coefficient measures the 2x overlap between the masked and predicted segmentations over the total number of pixels in two radiographs \cite{Radiopaedia}:

\[ DSC = \frac{2|A \cap B|}{|A| + |B|} \]

2. \textbf{Intersection over Union (IoU)}: IoU, also known as the Jaccard Index, is another fundamental metric for segmentation evaluation. Similar, to DSC it also quantifies the degree of overlap between the predicted region and the ground truth. However, it calculates the ratio of their intersecting area to their union. IoU provides a clear picture of how well the model's prediction aligns with the actual region.

IoU offers a straightforward interpretation: a higher IoU score indicates better model performance. It is particularly valuable in scenarios where accurate boundary delineation is essential. For instance, in medical imaging, precise segmentation is critical for diagnosing and treating various conditions. \cite{yu2021learning}:

\[ IoU = \frac{\text{Area of Overlap}}{\text{Area of Union}} \]

In summary, DSC and IoU are indispensable metrics and are most popular in the evaluation of segmentation models. They allow a comprehensive assessment of model accuracy by considering the balance between true positive rate, false positive rate, and false negative rate. These metrics enable researchers and practitioners in various fields, to ensure that segmentation models meet the rigorous standards required for accurate and reliable results.

\section{Results and Discussion}
In our study, we rigorously assessed nine distinct deep-learning-based image segmentation models. We utilized two key metrics to comprehensively evaluate their performance: Dice Similarity Coefficient (DSC), and Intersection over Union (IoU). Initially, we experimented with various loss functions for Mental Foramen (MF) segmentation, including Binary Cross Entropy, Focal Loss, and Dice Coefficient Loss. After a thorough evaluation, we determined that the Dice loss function was the most effective for the segmentation task. Therefore, we used Dice loss as the primary evaluation criterion for assessing the models.

To the best of our knowledge, our study stands as the first to employ both round-shaped and square-shaped masks as ground truth for training deep neural networks in the detection and segmentation of MF (Mental Foramen). The results obtained from the 5-fold cross-validation are presented in Table \ref{tab:combined_round_mask} for round-shaped masks and Table \ref{tab:combined_square_mask} for square-shaped masks.

\begin{table}[h]
    \centering
    \caption{Image segmentation on validation data with 5-fold CV and test data using the round-shaped mask for evaluation.}
    \begin{tabular}{|c | c c | c c|}
        \hline
        Models & \multicolumn{2}{c|}{5-fold CV Validation Data} & \multicolumn{2}{c|}{Testing Data} \\ 
        \cline{2-5}
        & DSC & IoU & DSC & IoU \\
        \hline\hline
        U--Net & \textbf{0.9782} & 0.6638 & \textbf{0.7902} & \textbf{0.6776} \\
        U--Net++ & 0.8753 & 0.7137 & 0.4963 & 0.6074 \\
        ResU-Net & 0.0124 & 0.4995 & 0.0866 & 0.5433 \\
        ResU-Net++ & 0.9152 & \textbf{0.7205} & 0.6758 & 0.6763 \\
        U--Net Attention & 0.9186 & 0.7272 & 0.6835 & 0.6716 \\
        FPN ResNet18 & 0.9057 & 0.6861 & 0.5097 & 0.5478 \\
        FPN InceptionV3 & 0.9112 & 0.7012 & 0.6624 & 0.6468 \\
        LinkNet ResNet18 & 0.8757 & 0.6382 & 0.6008 & 0.5480 \\
        LinkNet InceptionV3 & 0.9116 & 0.6691 & 0.6700 & 0.6180 \\
        \hline
    \end{tabular}
    \label{tab:combined_round_mask}
\end{table}

\begin{table}[h!]
    \centering
    \caption{Image segmentation on validation data with 5-fold CV and test data using the square mask for evaluation.}
    \begin{tabular}{|c | c c | c c|}
        \hline
        Models & \multicolumn{2}{c|}{5-fold CV Validation Data} & \multicolumn{2}{c|}{Testing Data} \\ 
        \cline{2-5}
        & DSC & IoU & DSC & IoU \\
        \hline\hline
        U--Net & \textbf{0.9884} & 0.8376 & \textbf{0.7996} & \textbf{0.7669 }\\
        U--Net++ & 0.9561 & 0.8973 & 0.7714 & 0.7638 \\
        ResU-Net & 0.7002 & 0.4818 & 0.5380 & 0.4864 \\
        ResU-Net++ & 0.9738 & \textbf{0.8974} & 0.7608 & 0.7618 \\
        U--Net Attention & 0.9866 & 0.9008 & 0.7985 & 0.7859 \\
        FPN ResNet18 & 0.9391 & 0.6303 & 0.7759 & 0.4865 \\
        FPN InceptionV3 & 0.9284 & 0.6024 & 0.7569 & 0.6114 \\
        LinkNet ResNet18 & 0.8863 & 0.4825 & 0.6959 & 0.4865 \\
        LinkNet InceptionV3 & 0.9252 & 0.4825 & 0.7486 & 0.4865 \\
        \hline
    \end{tabular}
    \label{tab:combined_square_mask}
\end{table}

Our findings revealed that the U-Net architecture consistently performed exceptionally well for both round-shaped and square-shaped masks on the validation data using the 5-fold cross-validation. Furthermore, we explored transfer learning using Inception V3 and ResNet 18 as backbones; however, we observed no significant differences in the DSC (Dice Similarity Coefficient) and IoU (Intersection over Union) metrics. For round-shaped masks in the (5-fold CV) validation data, the best performance metrics were as follows: DSC of 97.82\% with U-Net, IoU of 72.05\% with ResU-Net++. Similarly, for square-shaped masks in the (5-fold CV) validation data, the top-performing results were as follows: DSC of 98.84\% with U-Net, IoU of 89.74\% with ResU-Net++. On the test set, U-Net once again exhibited superior performance for both round-shaped and square-shaped masks, as detailed in Table \ref{tab:circle_test_table} and Table \ref{tab:square_test_table}. The best test performance metrics for round-shaped and square-shaped masks are as follows: 

1) Round-shaped mask: 
   - Dice Similarity Coefficient (DSC): 79.02\% with U-Net
   - Intersection over Union (IoU): 67.76\% with U-Net

2) Square-shaped mask: 
   - Dice Similarity Coefficient (DSC): 79.96\% with U-Net
   - Intersection over Union (IoU): 76.69\% with U-Net

\begin{figure}[h!]
\centering
\includegraphics[width=12cm, height=8cm]{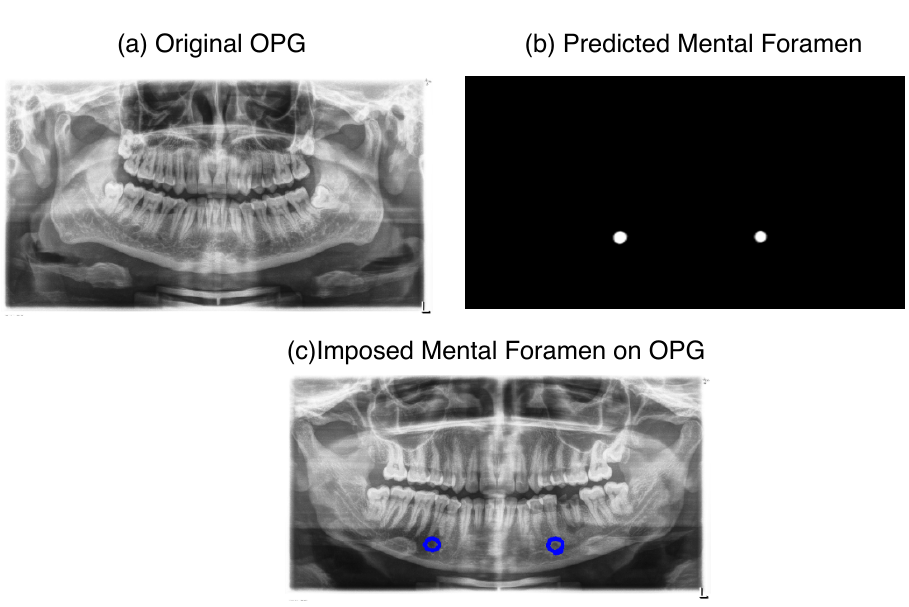}
\caption{Segmentation and detection of mental foramen (MF) on round-shaped mask: (a) original OPG; (b) predicted MF from original OPG given the image a; and (c) imposed predicted MF on original OPG given image a.}
\label{MF_round}
\end{figure}

\begin{figure}[h!]
\centering
\includegraphics[width=12cm, height=8cm]{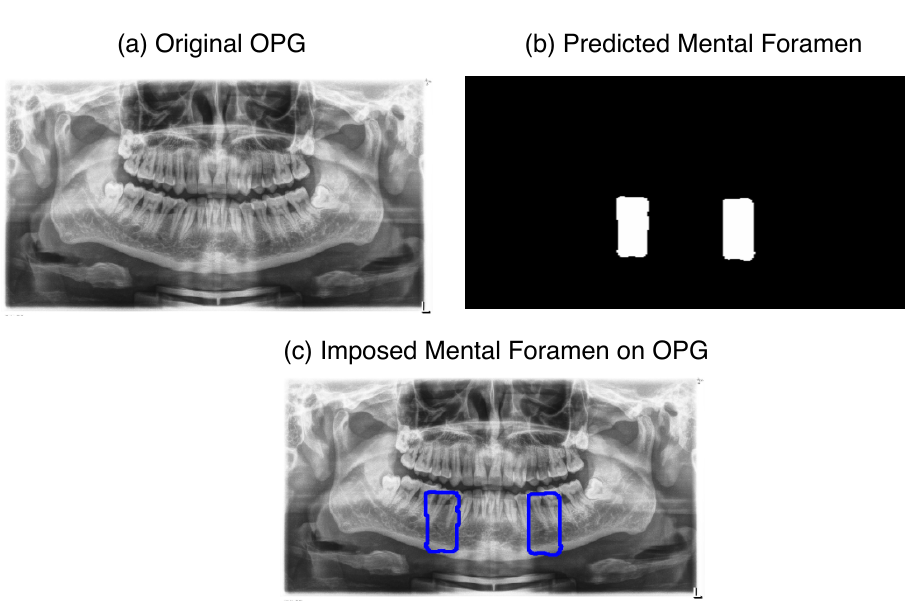}
\caption{Segmentation and detection of mental foramen (MF) on square-shaped mask: (a) original OPG; (b) predicted MF from original OPG given the image a; and (c) imposed predicted MF on original OPG given image a.}
\label{MF_square}
\end{figure}

In Figure \ref{MF_round}, we present the successful detection and segmentation of MF using the round-shaped mask. Figure \ref{MF_round} (a) showcases a randomly selected original OPG (Orthopantomogram) image from the test dataset, while Figure \ref{MF_round} (b) reveals the MF prediction generated by the U-Net model. In Figure \ref{MF_round} (c), we superimpose the segmented and detected MF onto the original image. Likewise, in Figure \ref{MF_square}, we exhibit the detection and segmentation of MF employing the square-shaped mask. Figure \ref{MF_square} (a) features a randomly chosen original OPG from the test dataset, while Figure \ref{MF_square} (b) displays the U-Net model's predicted MF. In Figure \ref{MF_square} (c), we provide a visual representation of the segmented and detected MF overlaid on the original image. Both Figure \ref{MF_round} and Figure \ref{MF_square} eloquently demonstrate the deep neural networks' remarkable ability to accurately detect and segment MF, closely aligning with the ground truth data.

\begin{figure}[h!]
\centering
\includegraphics[width=12cm, height=8cm]{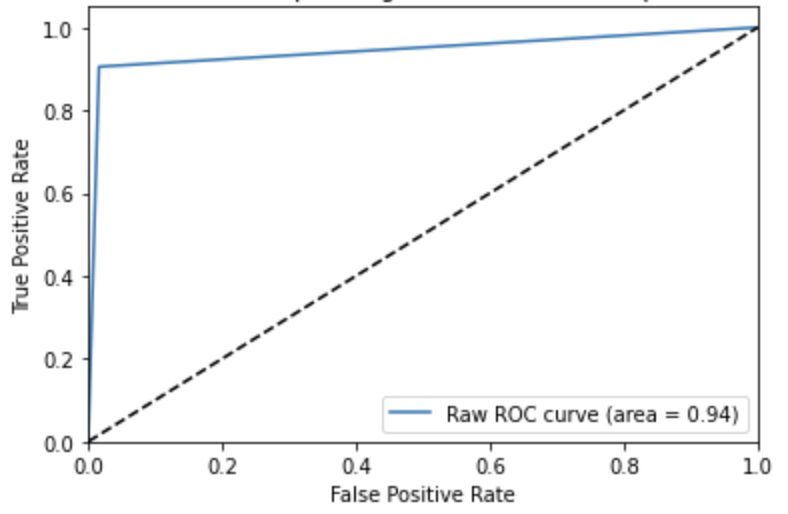}
\caption{ROC curve for the best model: U-net}
\label{roc}
\end{figure}

\section{Conclusion}
In the realm of medical imaging, the detection and segmentation of Mental Foramen (MF) within orthopantomography (OPG) radiographs hold paramount importance. Extensive research endeavors have marked the trajectory of this field, with recent years witnessing a notable shift toward deep learning techniques, which have exhibited great promise.

Traditionally, the task of detecting and segmenting MFs in X-ray images centred around a singular MF. However, our research has pioneered a more comprehensive approach by harnessing both MFs for model training and testing. The deep learning algorithm we have developed for this biomedical application has yielded impressive and efficient results. Among the nine models subjected to evaluation, the U-Net architecture has emerged as the standout performer, showcasing exceptional performance with both round-shaped and square-shaped masks. U-Net now stands as a frontrunner in the dental science domain, revolutionizing the landscape.

Our research incorporates several essential elements to bolster the model's detection capabilities, including the utilization of the Dice Similarity Coefficient (DSC), Intersection over Union (IoU), transfer learning via backbone models, and the application of cross-validation techniques. The transfer learning models were initialized with the same pre-trained weights, ensuring consistency and robustness.

The significance of automating MF detection and segmentation has surged, driven by the ever-growing volume of patient medical data, including X-ray radiographs, and the limited availability of specialized medical professionals. By harnessing the power of deep learning and advanced methodologies, our research makes a substantial contribution to addressing the critical need for accurate and efficient MF detection and segmentation in the field of dental science. It is poised to facilitate more accessible and precise dental diagnostics, ultimately enhancing patient care and outcomes.

\bibliographystyle{unsrt}  
\bibliography{references}

\end{document}